\input{aipcheck}

\documentclass[
    ,final            
  ]
  {aipproc}

\layoutstyle{6x9}

\begin{document}

\title{The Intra-Cluster Medium}

\author{Silvano Molendi}{
  address={Istituto di Astrofisica Spaziale e Fisica Cosmica, Sezione di Milano,
Via Bassini 15, I-20133 Milano, ITALY}
}

\begin{abstract}

The Intra-Cluster Medium (ICM) is a rarefied, hot, highly ionized,
metal rich, weakly magnetized plasma. In these proceeding, after
having reviewed some basic ICM properties, I discuss recent
results obtained with the BeppoSAX, XMM-Newton and Chandra
satellites. These results are  summarized in the following five
points. 1) Currently available hard X-ray data does not allow us
to constrain B fields in radio halos, the advent of hard X-ray
telescopes in a few years may change the situation substantially.
2) There is mounting evidence that temperature profiles of
clusters at large radii decline; however investigation of the
outermost regions will have to await a new generation of yet
unplanned but technologically feasible experiments. 3) The ICM is
polluted with metals, the enrichment has probably occurred early
on in the cluster's life. The abundance excess observed at the
center of CC clusters is due to the giant elliptical always found
in these systems. 4) Chandra and XMM-Newton observations of
relaxed clusters have falsified the previously accepted cooling
flow model, heating mechanisms that may offset the cooling are
actively being sought. 5) The superb angular resolution of Chandra
is allowing us to trace a previously unknown phenomenon intimately
related to the formation of galaxy clusters and of their cores.

\end{abstract}

\maketitle


\section{Setting the context}
In this section I shall review some of the basic properties of the
ICM, a more detailed discussion may be found in the excellent book
``X-ray emission from clusters of galaxies'' \cite{sarazin}. The
ICM is a tenuous plasma, typical densities range between
10$^{-4}$cm$^{-3}$ in the outer regions of clusters and a few
10$^{-2}$cm$^{-3}$ in the core of more relaxed systems (e.g.
\cite{mohr}). These are very low densities; the tail lobes of the
earth magnetosphere, which are considered the "best-vacuum" in the
earth's vicinity, have densities of the order of
10$^{-2}$cm$^{-3}$. The ICM is hot, temperatures are in the range
of 10$^7$ to 10$^8$ K (1-10 keV), it is also highly ionized; light
elements such as H and  He are completely ionized, heavier
elements are partially ionized. The ICM is chemically enriched,
with  heavy elements such as O, Mg, Si, Ar, Ca, Fe  and Ni, in
almost solar proportions. One issue, which is often overlooked, is
that the ICM is a magnetized plasma, with B fields in the range
between 0.1$\mu G$ and a few $\mu G$ (e.g. \cite{govoni} and
references therein). These values are the smallest found in any
astrophysical context.

%

The ICM emits radiation at X-ray wavelengths,
this emission is resolved by imaging X-ray experiments, current instrumentation
traces emission out to 1-2 Mpc from the core.
The ICM cools by emitting radiation, the cooling timescale, $t_{cool}$, may be
estimated from the ratio $u/\epsilon$ where $u$ is the energy density and
$\epsilon$ the emissivity,

$$ t_{cool} \sim 8.5 \times 10^{10} yr \Bigl ({n_p \over 10^{-3} cm^{-3}} \Bigr)^{-1}
   \Bigl ({T \over 10^{8} K} \Bigr)^{1/2}, $$
where $n_p$ is the proton  density and $T$ is the plasma temperature.
 Except for the innermost regions where  $n_p$ is high,
gas cools on timescales larger than the Hubble time (i.e. the age
of the Universe, $\sim$ 14Gyr) which is taken as an estimate of
the age of galaxy clusters. Thus, at least to first approximation,
we may consider the ICM as a stationary ball of hot plasma.
Because of the very slow rate at which the plasma is losing energy
by emitting X-rays no major on-going heating of the gas is
necessary. The ultimate origin of the bulk of the thermal energy
of the ICM is the gravitational energy lost by the plasma as it
falls into the cluster's potential well. Evidence of non
gravitational forms of heating have been found over the last
decade (e.g. \cite{ponman}), however the contribution of such
heating is thought to be more important in the less massive
systems, i.e. galaxy groups, than in galaxy clusters. The
temperature of the ICM is related to the depth of the potential
well and to the total mass of the cluster.

As already pointed out, the ICM is highly ionized, under such
conditions  Coulomb collisions are the dominant forms of
interaction in the plasma. Under the reasonable assumption that
electrons and ions are in thermal equilibrium amongst themselves
and each other, the mean free path for both species is:

$$\lambda_e \sim \lambda_i \sim 23 kpc \Bigl ({n_p \over 10^{-3} cm^{-3}} \Bigr)^{-1}
   \Bigl ({T \over 10^{8} K} \Bigr)^{2}. $$
The mean free path is much smaller than the cluster itself and the
ICM may be treated as a fluid satisfying hydrodynamical equations.

The timescale over which a sound wave crosses the cluster, $t_s$ is:

$$t_s \sim 6.6 \times 10^8 yr \Bigl ({D_c \over Mpc} \Bigr)
                        \Bigl ({T \over 10^{8} K} \Bigr)^{2}, $$
where $D_c$ is a measure of the cluster size.
Since $t_s$ is typically smaller than the cooling timescale and of the other 
timescales regulating the behavior of the cluster, the ICM may be assumed to be, 
at least to first approximation, in hydrostatic equilibrium.

The pressure associated to the ICM magnetic field is typically
much smaller than the thermal pressure of ions and electrons, thus
the magnetic field is not expected to drive the dynamics of ICM.
This however does not mean that the magnetic field has no effect
on the ICM, indeed the electron gyration radius, $r_e$ which is:

$$r_e \sim 7.0 \times 10^{-14} kpc \Bigl ({B \over \mu G} \Bigr)^{-1}
                        \Bigl ({T \over 10^{8} K} \Bigr)^{1/2}, $$
is many orders of magnitude smaller than the electron  mean free
path. This implies that electrons, as well ions for which a
similar results holds, will be forced to spiral along magnetic
field lines. This is expected to affect some of the properties of
the ICM such as its ability to conduct heat.

\section{Deriving physical quantities}
In this section I shall briefly review the methods employed to
measure some of the fundamental quantities characterizing the ICM.
The ICM density $n$, temperature $T$, and metalicity $Z$ are
usually recovered from X-ray measurements. The continuum emission
is dominated by thermal bremsstrahlung, whose emissivity,
$\epsilon_\nu$, may be expressed as $\epsilon_\nu \propto n_e^2
T^{-1/2} \exp(-h\nu/kT)$; the temperature is determined from the
position of the exponential cut-off in the X-ray spectrum. An
alternative method, which is sometimes employed, is to measure the
ratio of emission lines produced by the same atomic species (e.g.
\cite{gasta}). Since the plasma is in ionization equilibrium the
line ratio can be used to estimate the ionization state of the
specific atomic species and consequently the temperature of the
plasma. The gas density is estimated from the X-ray luminosity:
$$L(\nu_1,\nu_2) = \int_{\nu_1}^{\nu_2} d\nu \int_{V} dV \epsilon_\nu
\propto <n_e^2> T^{1/2}, $$ where [$\nu_1,\nu_2$] and $V$ are
respectively the frequency range and the volume over which
$\epsilon_\nu$ is integrated. The estimate requires some
assumptions on the distribution of the gas within the cluster.
Alternatively, if spatially resolved measurements are available,
deprojection techniques (see \cite{pizzo} and references therein)
may be used to derive density profiles. The metal abundances for a
given species is recovered from measurements of the equivalent
width of emission lines radiated by that specie. Indeed the
equivalent width of a line scales linearly with the metal
abundance of the element producing it.

Independent constraints on the physical quantities describing the
ICM may be obtained from measurements performed at the other end
of the electromagnetic spectrum through the Sunyaev \& Zeldovich
(hereafter SZ) effect. The SZ effect may be described as the
distortion of the cosmic microwave background spectrum by the
electrons in the ICM through Inverse Compton scattering. A key
parameter describing the strength of the distortions is the
Compton $y$ parameter defined as: $y\equiv \int (kT/ m_ec^2)
\sigma_T n_e dl$, where $\sigma_T $ is the Thompson cross section
and the integration is on the line of sight. At the present time
SZ measurements are much less sensitive than measurements
performed at X-ray wavelengths, however the situation may change
radically with the coming on line of new instrumentation over the
next decade \cite{carlstrom}. It is worth noting that while the
bremsstrahlung emissivity scales like $n_e^2$ the $y$ parameter
scales like $n_e$, thus the SZ effect may, in the not too distant
future, allow us to map the low density ICM in the outer regions
of galaxy clusters unaccessible to current X-ray instrumentation.
Perhaps of even greater importance is the fact that the $y$
parameter may be viewed as a sort of line of sight integrated
pressure ($y$ scales as the product of temperature and density).
This of course means that spatially resolved SZ measurements will
be sensitive to pressure gradients and discontinuities in much the
same way that spatially resolved X-ray measurements are sensitive
to $n_e^2$ gradients and discontinuities. Thus spatially resolved
SZ measurements may turn out to be the most effective way to
detect and characterize shocks (pressure discontinuities) which we
know must be present in the ICM  (e.g. \cite{ricker}) but that
have so far been very hard to find at X-ray wavelengths, probably
because they mostly occur in the low density outer regions of
clusters. The characterization of the thermodynamical properties
of the ICM in the outer regions of clusters is, in my view, a key
issue and I will come back to it in the next section.

The intensity of the magnetic field is generally estimated from
radio observations; the Faraday rotation measure, $RM$, of cluster
or background radio sources is combined with measurements of the
electron density to estimate the magnetic field of the plasma from
the relation: $ RM \propto \int n_e \vec{B} d\vec{l} $, where the
integration is extended over the line of sight. There are however
large uncertainties associated to these measurements. Firstly only
a few, not necessarily representative lines of sight are sampled
\cite{rudnick}; secondly the estimate of the magnetic field
intensity is critically dependent upon the scales over which the
field is ordered \cite{ensslin}.

\subsection{Hard tails}

There is an alternative way of estimating magnetic fields, at
least in some clusters. This method takes advantage of the fact
that some clusters feature diffuse synchrotron emission (radio
halos). In these cases, cosmic microwave background photons will
interact with the radio emitting electrons and be inverse Compton
scattered to high energies, where they can be detected. Since the
synchrotron luminosity, $L_{sync}$, and the Compton luminosity,
$L_{IC}$, scale respectively as $n_e U_B$ and $n_e U_{CMB}$, where
$n_e$ is the relativistic electron density and $U_B$ and $U_{CMB}$
are respectively the magnetic field energy density and Cosmic
Microwave background energy density, it is easy to see that:
$L_{IC} \propto L_{sync} U_{CMB} / U_B $. Thus joint radio and
X-ray measurements can lead to an independent and firmer
measurement of the magnetic field in clusters with radio halos.
First detection of a possible IC component has been achieved with
the PDS instrument on-board BeppoSAX in Coma \cite{ff99}, A2256
\cite{ff00} and A754 \cite{ff03}. In these objects the Authors
report detection of excess emission above the thermal component at
high energies (hard tails). The strongest detection was on Coma.
In this cluster  \cite{ff99}, using the method described above,
estimated a magnetic field of $\sim$ 0.14 $\mu G$.

\begin{figure}[h]
  \includegraphics[height=.50\textheight,angle=-90]{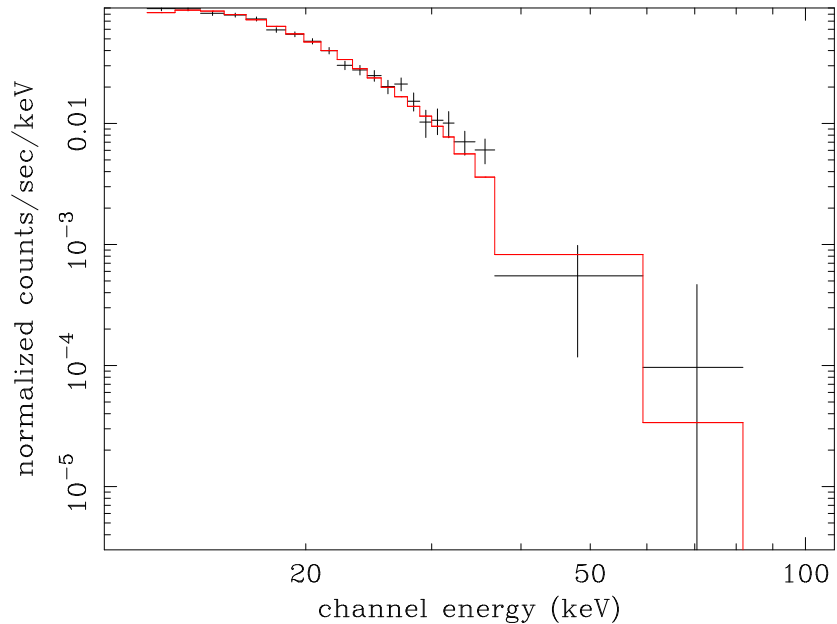}
  \caption{Coma spectrum measured with the PDS on board BeppoSAX, data is
from the second Coma observation (December 2000). The crosses are the data,
the continuous line is the best fit with a thermal model.}
\end{figure}

Recently \cite{ross}  have analyzed a new, longer and lower
background PDS observation of the Coma cluster finding no evidence
for a hard tail (see Fig. 1). Moreover, work in collaboration with
the PDS Hardware Team has lead to the identification of an error
in the reduction leading to the spectrum published in \cite{ff99},
once this is corrected for, the significance of the tail drops
below 3$\sigma$. \cite{ross} conclude that there is no compelling
evidence for a hard tail in the PDS Coma spectrum (see however
\cite{ff03bis} for a different point of view). In summary the
measurement of the IC component offers an elegant and potentially
powerful method to determine magnetic fields in clusters with
radio halos. However, to employ this method fruitfully, we need
substantially more sensitive hard X-ray instrumentation. A
breakthrough may come in the not too distant future when the first
hard X-ray telescopes will be flown.


\section{Temperature Profiles}

Spatially resolved measurements of the ICM temperature are
important for two main reasons: firstly they can be used to
determine the total mass of clusters through the hydrostatic
equilibrium equation (only about 20-30\% of the total mass in
clusters is in the form of visible matter, while the bulk of the
mass is in the form of non-baryonic dark matter); secondly they
provide important clues on the thermodynamic state of the ICM.

\begin{figure}[t]
 \includegraphics[height=.61\textheight,angle=-90]{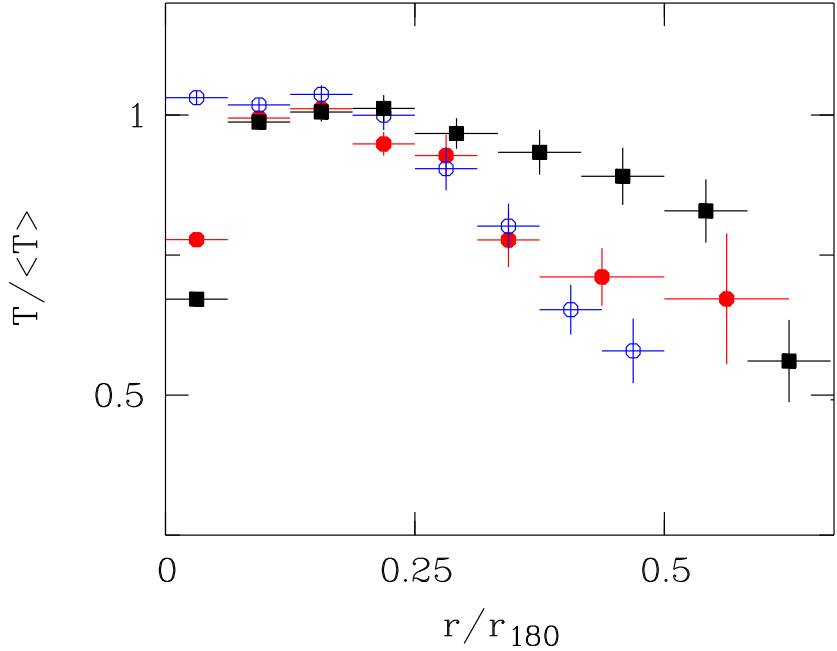}
 \caption{Mean temperature profile from  BeppoSAX \cite{dm02} (circles) and
XMM-Newton \cite{dg04} (squares) data. The BeppoSAX sample is
divided in cool core (filled circles) and non cool core (empty
circles) clusters. The radius is in units of $r_{180}$ and the
temperature in units of the cluster mean temperature.}
\end{figure}

In Fig. 2 I report the mean temperature profile for a sample of 23
galaxy clusters observed with BeppoSAX \cite{dm02}. The  sample is
divided in cool core (CC) and non cool core (NCC) clusters. As can
be seen from Fig. 2, beyond the core where CC clusters show a
decrement (this is indeed the reason why they are called Cool Core
clusters), the mean CC and NCC profiles are similar. Up to about
0.2 r$_{180}$  the profiles are flat, beyond this radius the
profiles decline. It should not go unmentioned that the existence
and extension of the outermost region has been the subject of a
lively debate \cite{mark98}, \cite{irwin99}, \cite{white00},
\cite{irwin00}. By fitting the temperature profiles from $r > 0.2
r_{200}$ with a power law of the form $T(r) \propto r^{-\mu}$ and
assuming that the density profile is described by a power law of
the form $n(r) \propto r^{-2}$, we find that the polytropic index,
$\gamma$,  for CC and NCC clusters is respectively 1.22$\pm$0.04
and 1.16$\pm$0.03. These values are about half way between the
isothermal, $\gamma =1$ and adiabatic $\gamma =5/3$ case. In Fig.
2 we also compare the mean BeppoSAX profile with a mean XMM-Newton
profile \cite{dg04}. The latter has been obtained by averaging the
recently published temperature profiles of 14 galaxy clusters in
the redshift range $0.1 < z <0.3$. 
As can be seen the XMM-Newton profile, albeit qualitatively similar, 
appears to be somewhat flatter than the Beppo-SAX one.
However the most
interesting point is that, contrary to pre-flight expectations the
XMM-Newton profiles do not extend significantly further out than
the BeppoSAX profiles. This is particularly unfortunate as the
regions around the virial radius ($r_{vir} \sim r_{180}$) are
expected to carry much information on the formation process of
galaxy clusters (e.g. \cite{tozzi00}). Indeed I am inclined to
believe that a solid observational characterization of these
regions would allow us to improve considerably our understanding
of the physics of galaxy clusters as a whole perhaps providing
important clues to the solution of outstanding open problems.
It is a matter of considerable concern that currently
planned missions will not be able to address this important issue
although, from a technological point of view, the problem of
constructing an experiment sensitive to low surface brightness
regions is not a particularly challenging one.

 \section{Abundance Profiles}

Clusters are the largest structures in the Universe to have
clearly decoupled from the Hubble flow, in principle the ICM could
be made of H and He only. The presence of heavy elements such as
Fe in proportions which are almost solar demonstrates that a
sizeable fraction of the ICM must have been processed in stars.
This establishes an important connection between the Galaxies and
the ICM in clusters. Global measurements indicate that the ICM in
clusters has a mean metalicity of $Z \sim 0.3 Z_\odot$ with a
rather small dispersion around this mean. \cite{arnaud92} have
shown that for a sample of rich clusters, there is a strong
correlation between the total Fe mass in the ICM and the
bolometric luminosity of early-type galaxies (E/S0). No
correlation is present with spiral galaxies  indicating that a
non-negligible fraction of the ICM likely originates in E/S0
galaxies. Another important piece of information is that there is
no evidence of variations of $Z$ with redshift out to $z \sim 1.2$
(\cite{tozzi03} and references therein), implying that the bulk of
the enrichment occurs at redshifts larger than $\sim 1.2$.

\begin{figure}
  \includegraphics[height=.61\textheight,angle=-90]{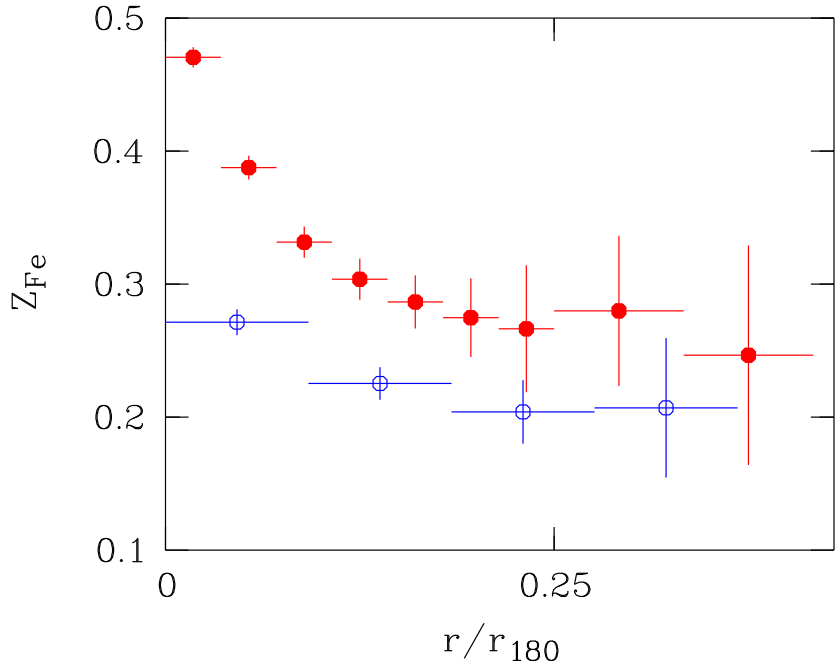}
  \caption{Mean Fe abundance profile from the BeppoSAX sample \cite{dm01}.
The sample is divided in cool core (filled circles) and non cool core
(empty circles) clusters.
The radius is in units of $r_{180}$ and the Fe abundance in solar
units.}
\end{figure}

From the analysis of a sample of 23 galaxy clusters observed with
BeppoSAX \cite{dm01} (see Fig. 3) we have found that while NCC
clusters show essentially flat abundance profiles, CC clusters
show evidence of an abundance excess in their core. Detailed
investigations \cite{fuka00}, \cite{dm01}, \cite{dg03} have shown
that the excess observed in the core of CC clusters (see Fig. 3)
is due to enrichment from the giant galaxy found at the center of
these systems. The amount of iron associated to the excess  ranges
from a few $10^9$ to a few $10^{10} M_\odot$ and is a relatively
low fraction of the total Fe mass contained in the cluster, less
than $\sim$10\%, \cite{dg03}.

\section{The new stuff}

In the last 3 years observations carried out with the latest
generation of X-ray satellites, i.e. Chandra and XMM-Newton, have
brought about some rather important changes in our understanding
of galaxy clusters. In the this section I will cover two topics
which have greatly benefited from the new X-ray observations.

\subsection{Cool Cores}

Since this topic is covered in  detail elsewhere in this volume
\cite{fab04}, I will limit myself to some elementary
considerations and then focus on a specific object that I find of
particular interest.

Chandra and XMM-Newton observations of relaxed clusters have shown
that the cool gas in their core is characterized by a minimum
temperature of 1-3 keV (e.g. \cite{peterson03}, \cite{molendi01}).
Moreover detailed analysis of some close by clusters (e.g. Virgo;
\cite{molendi02} and next subsection) have shown that the gas is
well described by one or two temperature models. Both these
observational findings are in clear violation of the previously
adopted cooling flow model. The fall of the standard cooling flow
models leaves a fundamental question open: what happens to the gas
that should be cooling on very short timescales? Two classes of
solutions have been proposed: in the first class \cite{fabian01}
the cooling gas is there but is somehow hidden from our view; in
the second class the gas is prevented from cooling below a minimum
temperature by some form of heating: various mechanisms have been
proposed, some have to do with heating from the ICM  outside the
cool core (e.g. \cite{narayan02}, \cite{fabian02},
\cite{fujita03}), some invoke heating from the Active Galactic
Nucleus commonly found in the core of these systems (e.g.
\cite{ruszkowski02}, \cite{churazov02}).
Although much work has been done, most heating mechanisms
that have been identified so far seem to have rather obvious
flaws. Further exploitation and digestion of Chandra and XMM-Newton data
as well new observations with the soon to be launched ASTRO-E2 satellite may
hold the key to  the cool core puzzle.

\subsubsection{Virgo/M87}

The core of the Virgo cluster, hosting the giant elliptical galaxy
M87, is very close to us. At the distance of M87 1 arcminute
corresponds to about 5 kpc and very small structures are resolved.
Detailed analysis of the Chandra and XMM-Newton  data  show that
most regions in the core of Virgo are adequately described  by one
temperature models \cite{molendi02}, in a few regions, co-spatial
with intense radio emission \cite{owen00}, two temperature models
provide a much better description of the data. In these regions we
find that most of the gas has a temperature very similar to that
of nearby one temperature regions (hot component) and a small
fraction has a smaller temperature (cool component). The ratio of
the plasma density of the cool, $n_{cool}$, and hot, $n_{hot}$,
component may be expressed as: $n_{cool}/n_{hot} \sim ( EI_{cool}
V_{hot})^{1/2} / (EI_{hot} V_{cool})^{1/2}$, where $EI_{cool}$ and
$V_{cool}$ ($EI_{hot}$ and $V_{hot}$) are respectively the
emission integral and the volume associated to the cool (hot)
component. Assuming pressure equilibrium between hot and cool
components, $n_{hot} kT_{hot} = n_{cool} kT_{cool}$, it is easy to
show that $ V_{cool}/V_{hot} \sim EI_{cool}/ EI_{hot} (KT_{cool} /
KT_{hot})^2 $. Typical values of $V_{cool}/V_{hot}$ range between
$10^{-3}$ and $10^{-2}$, more precise calculations based on actual
deprojection of the data are in agreement within a factor of 2.
This implies that the size of individual cool structures is much
smaller than the size of the regions from which spectra have been
extracted. From images accumulated in two different energy bands,
the first sensitive to the cool component and the second to the
hot component, it is possible to derive $EI_{cool} / EI_{hot}$
images. This has been done both with XMM-Newton EPIC and Chandra
ACIS-S data. The images show that $EI_{cool} / EI_{hot}$ is almost
everywhere smaller than 0.4 implying that $V_{cool} / V_{hot}$ is
smaller than $\sim$ 0.2. The cool blobs are not resolved by
XMM-Newton or Chandra, implying that their typical size must be
smaller than $\sim $ 300pc. Since the timescale for conduction to
operate on such small scales, 10$^5$ yr, is much smaller than
other timescales presiding over the behavior of the plasma we
conclude that conduction must be heavily suppressed. For many
years it has been recognized that conduction might be suppressed
in the ICM and magnetic fields have been invoked as a possible
means to achieve suppression, it is therefore of considerable
interest that in the case of M87 the cool blobs, where conduction
is suppressed, are co-spatial with  radio lobes, where magnetic
fields are present.

\subsection{Cold Fronts}

One of the first Chandra observations was performed on the galaxy
clusters A2142 \cite{mark00}. The image revealed two sharp
bow-shaped, shock like, surface brightness edges. However these
are not shocks: shocks produce compression and heating, if these
structures were shocks, one would expect that the denser gas would
also be hotter, on the contrary the data reveals that the denser
gas is cooler. The discontinuities observed by Chandra are a new
phenomenon that has been termed ``cold front''. Cold fronts have
now been observed in the core of many  CC clusters, e.g. A1795
\cite{mark01}, Centaurus \cite{sanders02}, 2A 0335+096
\cite{mazzotta03}, and in merging clusters, e.g. A2142
\cite{mark00} and A3667 \cite{vikhlinin01}. In the latter case the
edges mark dense subcluster cores that have survived a merger,
while in the former they provide evidence of gas motions possibly
associated to past mergers \cite{mark01} or to the relative motion
of the cD galaxy with respect to the ICM \cite{fabian01b}. The
edges of cold fronts are very sharp, in some cases (e.g. A2142
\cite{ettori00}, A3667 \cite{vikhlinin01b}), it has be shown that
the surface brightness drops over scales smaller than the Coulomb
mean free path, indicating that heat conduction must be suppressed
across the discontinuity.

Recently \cite{mark03} has reported that
in a sample of 33 CC clusters density edges are seen in 19 (60\%).
Since typical velocities of the gas in the CC systems are in the order of  1/2
of the sound speed the associated energy will be about 1/4 of the thermal energy
of the plasma.
Of some interest is also the fact that gravitational mass estimates of cool
cluster cores based on the hydrostatic equilibrium equation must frequently
 be affected by systematic errors of the order of various tens of percent.

In the case of merging clusters the presence of  subcluster cores
that have survived a merger indicates that the virialization
process in these clusters is far from being complete. Indeed it is
quite likely that these substructures, or parts of them, will
never be completely dissolved within the ICM and that, after
having lost most of their kinetic energy, they will fall to the
bottom of the cluster potential well where they will eventually
contribute to the formation of a cool core.


\begin{theacknowledgments}
It is a pleasure to acknowledge my collaborators Fabio Gastaldello, Fabio
Pizzolato, Andrea De Luca, Simona Ghizzardi, Chiara Rossetti and Sabrina De Grandi.

\end{theacknowledgments}


\bibliographystyle{aipprocl} 

\end{document}